\documentclass[a4paper,aps,twocolumn,amsfonts,notitlepage,superscriptaddress,nofootinbib]{revtex4-2}
\pdfoutput=1

\usepackage[utf8]{inputenc}

\usepackage{hyperref}

\usepackage{amsmath}
\usepackage{amssymb}
\usepackage{xcolor}
\usepackage{physics}
\usepackage{graphicx} 

\renewcommand{\theequation}{\arabic{equation}}

\usepackage{pgf} 

\usepackage{tabularx}

\newenvironment{eqaed}
    {\begin{equation}
    \begin{aligned}
    }
    { 
    \end{aligned}
    \end{equation}
    }

\newcommand{\commie}[1]{}

\begin{document}

\author{Ivano Basile}
\email{ivano.basile@umons.ac.be}
\affiliation{Service de Physique de l'Univers, Champs et Gravitation, Universit\'{e} de Mons, Place du Parc 20, 7000 Mons, Belgium}

\author{Alessia Platania} 
\email{aplatania@perimeterinstitute.ca}
\affiliation{Perimeter Institute for Theoretical Physics, 31 Caroline St.  N., Waterloo, ON N2L 2Y5, Canada}
\title{String Tension between de Sitter vacua and Curvature Corrections}

\begin{abstract}
Higher-derivative corrections to cosmological effective actions in string theory are largely constrained by T-duality, but have been computed hitherto only to the first few orders in the string scale~$\alpha'$. The functional renormalization group, in conjunction with the strong constraints imposed by T-duality, allows to derive cosmological effective actions to all orders in~$\alpha'$ while avoiding ``truncations'' of the theory space. We show that the resulting higher-derivative $\alpha'$-corrections forbid the existence of de Sitter vacua, at least in regimes where string-loop corrections can be neglected. Our findings thus support the no-de Sitter swampland conjecture in the presence of all-order effects in~$\alpha'$.
\end{abstract}

\maketitle


\textit{Introduction.}---
Effective actions in quantum field theories (QFTs) embody quantum fluctuations on all scales in a single functional of the degrees of freedom of the theory. Their knowledge is thus paramount to shed light on a variety of phenomena that could be compared with observations. Whenever a field theory remains weakly coupled at all scales, perturbative methods generally provide adequate computational control. However, in the case of gravity, all-order (and possibly non-perturbative) effects appear necessary to capture all of the relevant physics. 

In asymptotically safe gravity \cite{1976W,1979W,Souma:1999at,Bonanno:2020bil} these effects ought to arise from a non-Gaussian ultraviolet fixed point, whose existence would make the theory ``asymptotically safe", i.e. non-perturbatively renormalizable in the Wilsonian sense~\cite{1976W,1979W}. In (perturbative) string theory, these effects arise from a double expansion involving the string scale~$\alpha'$ and the string coupling $g_s$. Although in principle well-defined, the computation in general settings is highly involved, and all-order computations in the absence of supersymmetry can be carried out only in specific backgrounds, such as the linear dilaton background and the ones studied in~\cite{Amati:1988ww, Amati:1988sa}.

In contrast, phenomenologically relevant settings, such as de Sitter (dS) vacua for both early- and late-time cosmology, have proven much subtler, since the supersymmetry algebra cannot be (linearly) realized in dS, and their existence in string theory remains under detailed scrutiny~\cite{Kachru:2002gs, Kachru:2003aw, Kutasov:2015eba, Danielsson:2018ztv, Cicoli:2018kdo, Gautason:2018gln, Hamada:2019ack, Gautason:2019jwq, Dasgupta:2019gcd}. Some supersymmetry-breaking mechanisms appear to yield naturally other types of cosmologies~\cite{Dudas:2000ff}, whose implications for Cosmic Microwave Background anomalies have been explored in~\cite{Kitazawa:2014mca, Kitazawa:2015uda, Sagnotti:2015asa}, while the associated instabilities suggest the possibility of dS braneworlds~\cite{Basile:2020mpt}. At present, however, the consistency of a number of dS constructions is unsettled, and a deeper understanding could require the inclusion of all-order effects.

In an attempt to shed some light on whether dS vacua are allowed in the presence of all-order effects, in this Letter we employ functional renormalization group (FRG) techniques~\cite{Dupuis:2020fhh} to resum $\alpha'$ corrections and investigate the resulting string cosmologies. In the context of the FRG, the object of interest is the effective average action (EAA)~\cite{Wetterich:1992yh}, since its flow allows, at least in principle, to infer the existence of an asymptotically safe regime, granting non-perturbative renormalizability~\cite{1976W}, and/or to recover the exact quantum effective action via its RG flow in the IR. One of the major practical limitations of this approach is typically the need to truncate the theory space in order to feasibly solve the flow equations~\cite{Dupuis:2020fhh}. However, the remarkable constraints that string theory entails on its low-energy theory space, which arises integrating out its higher-spin massive modes, may partially circumvent this issue. Concretely, in cosmological settings T-duality constrains the mini-superspace\footnote{Despite its limitations, the mini-superspace approach has proven useful in quantum cosmology, as for instance in~\cite{PhysRevLett.86.5227, PhysRevLett.119.171301, PhysRevLett.100.201301, PhysRevLett.121.081302, PhysRevLett.96.141301}. In the present setting, it allows to exploit the classification of~\cite{Veneziano:1991ek, Meissner:1991zj, Meissner:1996sa, Hohm:2015doa} in order to obtain all-order curvature corrections~\cite{Basile:2021amb}.} effective action to a large extent~\cite{Veneziano:1991ek, Meissner:1991zj, Meissner:1996sa, Hohm:2015doa, Hohm:2019ccp, Hohm:2019jgu}\footnote{The swampland program also purports constraining the low-energy theory space from the high-energy vantage point of string theory~\cite{Palti:2019pca,Danielsson:2018ztv}, and could potentially further simplify FRG computations.}, and could therefore go hand in hand with FRG methods to capture all-order effects in $\alpha'$. In this Letter we provide a proof of principle of this idea, applying the FRG to resum $\alpha'$ corrections to all orders in cosmological backgrounds, neglecting $g_s$ string loop corrections. Based on two closed-form solutions to the (mini-superspace) flow equations, we show that dS vacua are not compatible with the resulting quantum-corrected cosmologies, thus providing support to the no-dS swampland conjecture~\cite{Obied:2018sgi, Ooguri:2018wrx, Garg:2018reu, Palti:2019pca} and highlighting a tension between de Sitter vacua and curvature corrections in string theory.


\textit{Effective actions and cosmology.}---
The existence of dS solutions is a relevant open question in quantum gravity, in particular within the context of string theory and the swampland program at large~\cite{Danielsson:2018ztv}. Whether they can be realized relies crucially on the form of the gravitational quantum effective action. In any metric approach to quantum gravity based on QFT (e.g., quantum unimodular gravity, asymptotically safe gravity, Ho\v{r}ava-Lifshitz gravity, etc.), if diffeormorphism invariance holds at high energies and no additional symmetries constraining the gravitational interaction arise, the Einstein-Hilbert term ought to be complemented by all possible curvature invariants. Deriving the effective action exactly, at least within a mini-superspace scheme~\cite{Hawking:1983hn, Hawking:1983hj, Vilenkin:1994rn, Ashtekar:2011ni, Bojowald:2015iga}, requires summing over all quantum fluctuations at level of a gravitational functional integral or, equivalently, solving the FRG equations~\cite{Dupuis:2020fhh}. While these computations are generally unfeasible in the untruncated space of diffeomorphism-invariant theories (even in mini-superspace), the additional symmetries of string theory make it an ideal candidate in which to determine cosmological effective actions with FRG methods~\cite{Basile:2021amb} and investigate the existence of dS solutions.

\begin{figure}
	\hspace{-0.4cm}\includegraphics[scale=0.4]{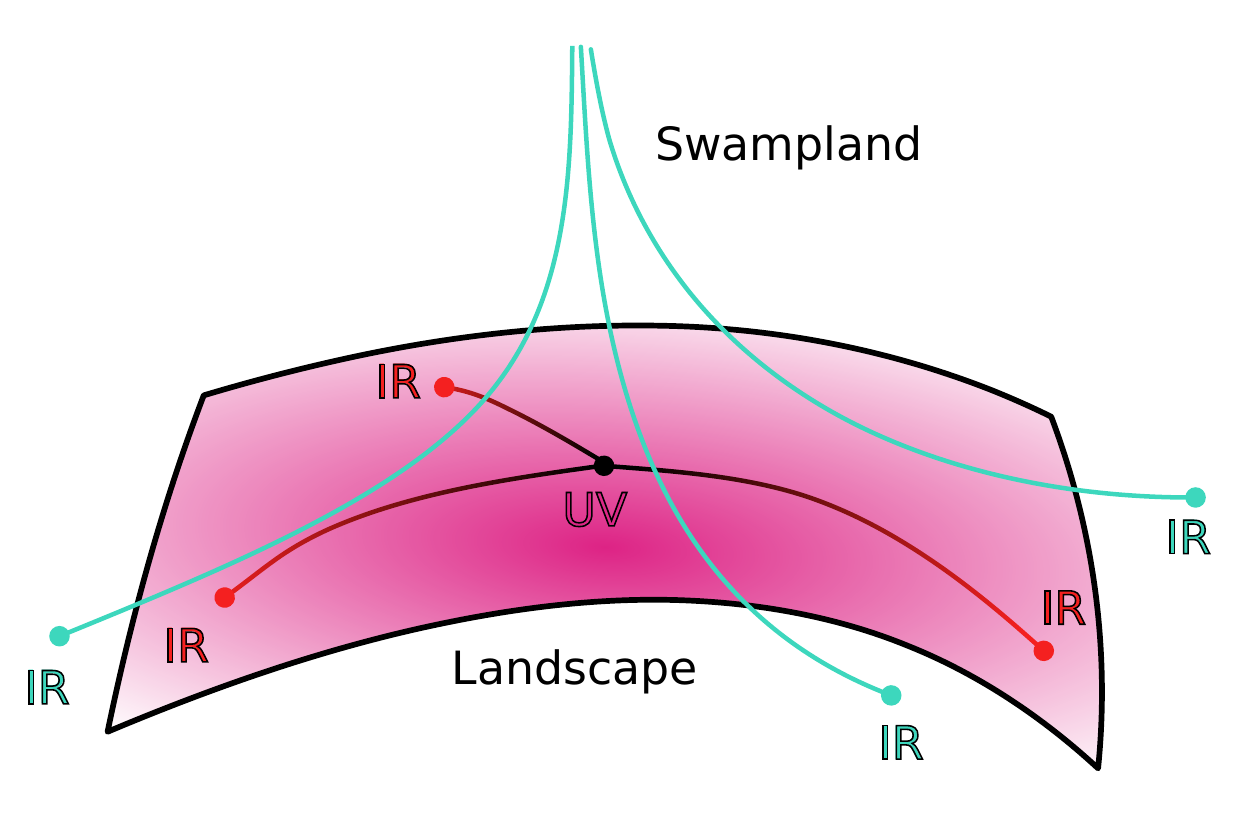}
	\caption{Pictorial representation of the ``landscape'' of low-energy theories which admit a UV completion as a UV critical surface. The trajectories off the critical surface (green curves) belong to the ``swampland'' of inconsistent theories. \label{Fig:swampland_landscape}}
\end{figure}

Gravitational effective actions in string theory can be derived, at least at leading order in the string coupling~$g_s$, from the worldsheet formulation and their form in cosmological backgrounds is strongly constrained by T-duality. Specifically, in $D = d + 1$ dimensions, T-duality emerges as an $O(d,d,\mathbb{R})$ symmetry on the low-energy degrees of freedom~\cite{Veneziano:1991ek, Meissner:1991zj, Meissner:1996sa}, namely the dilaton $\phi$, the metric $G$ and the Kalb-Ramond two-form $B_2$. Focusing on cosmological backgrounds of the type
\begin{eqaed}\label{eq:cosmological_ansatz}
	ds^2 & = - \, n^2(t) \, dt^2 + e^{2\sigma(t)} \, d\mathbf{x}^2 \, , \\
	B_2 & = 0 \, ,\qquad \Phi = \Phi(t) \, ,
\end{eqaed}
where we have defined $\Phi(t)$ in terms of $\phi$ as in~\cite{Meissner:1991zj, Meissner:1996sa, Hohm:2015doa, Hohm:2019ccp, Hohm:2019jgu}, the tree-level action reduces to 
\begin{eqaed}\label{eq:red_action}
	S_{\text{red}} = \frac{\text{Vol}_d}{16\pi G_{\text{N}}}\int dt \, \frac{1}{n} \, e^{-\Phi} \left( - \, \dot{\Phi}^2 + d \, \dot{\sigma}^2 \right) \, ,
\end{eqaed}
where $G_{\text{N}}$ is the $D$-dimensional Newton constant, $\text{Vol}_d$ is the volume of~$d$-dimensional spatial slices and $H=\dot{\sigma}$ is the Hubble parameter. All~$\alpha'$-corrections are encoded in the higher-derivative part of the Meissner-Hohm-Zwiebach effective action~\cite{Meissner:1991zj, Meissner:1996sa, Hohm:2015doa, Hohm:2019ccp, Hohm:2019jgu}. On cosmological backgrounds, T-duality  requires that the effective action be even in $H$. Specifically, it can be shown~\cite{Meissner:1991zj, Meissner:1996sa, Hohm:2015doa, Hohm:2019ccp, Hohm:2019jgu} that its all-order expression reads\footnote{Note that the effective action in  Eq.~\eqref{eq:alpha_corrections} is only valid for $g_s = e^\phi \ll 1$, whereby string-loop corrections are expected to be negligible. In this limit, T-duality is indeed realized as a continuous~$O(d,d,\mathbb{R})$ symmetry, whereas $g_s$-corrections generally break it to~$O(d,d,\mathbb{Z})$.}
\begin{eqaed}\label{eq:alpha_corrections}
	&S_{\text{HD}} \sim \frac{\text{Vol}_d}{16\pi G_{\text{N}}}\int dt \, \frac{e^{-\Phi}}{n} \times\\
	&\left[ - \, \dot{\Phi}^2 + 2 d \, n^2 \, \sum_{m=0}^\infty (-4)^m \, c_m \,\alpha'^{m-1} \left(\frac{H}{n}\right)^{2m} \right] \, .
\end{eqaed}
In string perturbation theory, obtaining the coefficients~$c_m$ is typically a daunting task, even for low-order coefficients. In this work, we shall attempt to overcome these issues and compute \textit{all} the $c_m$ at once via FRG techniques~\cite{Dupuis:2020fhh}. The latter allow one to determine the flow of of the EAA $\Gamma_k$ via the exact flow equation~\cite{Wetterich:1992yh,Morris:1993qb,Reuter:1996cp}
\begin{eqaed}\label{eq:wetterich}
k \partial_k \Gamma_k=\frac{1}{2}\,\mathrm{STr}\left\{ \left( \Gamma_k^{(2)}+\mathcal{R}_k \right)^{-1} k \partial_k \mathcal{R}_k \right\} \, .
\end{eqaed}
The super-trace on the right-hand-side denotes a sum over internal indices and an integral over continuum spacetime coordinates or momenta\footnote{Let us remark that functional traces ought to be defined via the inner product induced by the kinetic terms of the classical action.}. The function $\mathcal{R}_k$ is a regulator which implements the Wilsonian shell-by-shell integration of fluctuating modes: at a scale $k$, all quantum fluctuations with momenta $p^2>k^2$ are integrated out and the quantum effective action ought to be recovered in the IR limit $k\to0$.  To wit, the regulator $\mathcal{R}_k$ has to vanish as $k\to0$, where indeed $\Gamma_{k\to0}$ reduces to the standard quantum effective action. On the other hand, a UV-complete theory ought to flow to a fixed point as $k \to \infty$, while trajectories off the UV critical surface correspond to theories which are inconsistent, and thus belong to the ``swampland'', as shown in Fig.~\ref{Fig:swampland_landscape}. Let us also remark that non-locality, which is an expected feature of quantum gravity, can be treated within the FRG (see~\cite{Knorr:2019atm} for examples in the context of asymptotically safe gravity). Specifically, the formalism could in principle account for UV/IR mixing by constraining the IR physics via non-trivial requirements stemming from UV completeness.

Since T-duality imposes strong constraints on the form of the effective action in cosmological backgrounds, the flow of a scale-dependent Meissner-Hohm-Zwiebach EAA can be computed analytically, avoiding truncations of the theory space\footnote{Although T-duality constrains the effective action, the presence of a regulator could induce deviations from the hypersurface of T-duality-invariant theories. Nevertheless, since T-duality ought to re-emerge in the limit $k\to0$~\cite{Meissner:1996sa, Hohm:2015doa, Hohm:2019ccp, Hohm:2019jgu}, we expect these deviations to be small and we neglect T-duality-breaking terms.}. To this end, the first step is to promote all couplings in Eq.~\eqref{eq:alpha_corrections} to scale-dependent quantities
\begin{eqaed}
c_m \to c_m(k) \, , \qquad G_\text{N} \to G_k \equiv g_k \, \frac{k^{1-d}}{16\pi} \, ,
\end{eqaed}
where the factor of $16\pi$ is introduced for convenience. All $\alpha'$-corrections are thus conveniently encoded in an even, dimensionless, scale-dependent function $F_k$, according to
\begin{eqaed}\label{eq:EAA_ansatz}
\Gamma_k = \text{Vol}_d \int \hspace{-0.1cm} dt \, \frac{ e^{-\Phi}}{n}  \left[ - \, \frac{k^{d-1}}{g_k} \,  \dot{\Phi}^2 + k^{d+1} \, n^2 \, F_k\left(\frac{H^2}{n^2}\right) \right] .
\end{eqaed}
In what follows we shall gauge-fix $n = 1$, so that ghosts trivially decouple. 

Given a solution $\Gamma_k$, one can send~$k\to0$ and extract the corresponding coefficients $c_m\equiv c_m(0)$.
This procedure would then yield $\alpha'$-corrections for the Meissner-Hohm-Zwiebach action~\cite{Meissner:1991zj, Meissner:1996sa, Hohm:2015doa, Hohm:2019ccp, Hohm:2019jgu} to all orders, a task that would appear at present out of reach in perturbative string theory. In addition, the effective action~$\Gamma_0$ gives access to a number of observable quantities and allows, at least in principle, to compute solutions to the quantum-corrected field equations and address a number of interesting challenges, such as the puzzling apparent lack of dS solutions~\cite{Obied:2018sgi, Ooguri:2018wrx, Garg:2018reu}.


\textit{Existence of de Sitter vacua.}---
Resumming $\alpha'$-corrections to all orders entails solving Eq.~\eqref{eq:wetterich}
for the scale-dependent Meissner-Hohm-Zwiebach effective action in Eq.~\eqref{eq:EAA_ansatz}. To this end, one has to derive the beta functions for the coupling $g_k$ and the function $F_k$. This  amounts to replacing the the scale-dependent Meissner-Hohm-Zwiebach effective action of Eq.~\eqref{eq:EAA_ansatz} in Eq.~\eqref{eq:wetterich}. 
The flow equations for $g_k$ and $F_k(H^2)$ have been derived in \cite{Basile:2021amb}.
They admit a \textit{particular solution for any $d$} given by
\begin{equation}
g_{k}=\frac{g_{0}g_{\ast}}{g_{\ast}k_{0}^{d-1}+g_{0}(k^{d-1}-k_{0}^{d-1})} \, k^{d-1}\label{eq:runningg} \, ,
\end{equation}
\begin{equation}
F_k(H^{2})=k^{-d-1}\left(c_{0}+c_{1}\sqrt{H^{2}}\right) \, ,
\end{equation}
where $c_{0}$, $c_{1}$ and $g_0$ are integration constants, $k_{0}$ is a reference scale, and
$g_{\ast}=2^{d}d(d-1)\pi^{1+\frac{d}{2}}\Gamma\left(\frac{d}{2}\right)$ is the position of the non-Gaussian UV fixed point for $g_k$. Interestingly, despite T-duality, the running of the Newton coupling in Eq.~\eqref{eq:runningg} matches the one found in asymptotically safe gravity~\cite{Bonanno:2000ep}. This suggestive result is non-trivially consistent with our scenario, where the string effective action arises in the IR from a UV fixed-point action that is closely approached by the RG trajectory of string theory. In particular, this resonates with the intriguing possibility that asymptotically safe gravity sits in a corner of the theory space of string theory~\cite{deAlwis:2019aud}, thus realizing a scenario similar to that advocated by Weinberg in~\cite{Weinberg:2021exr}\footnote{See also~\cite{Basile:2021krr} for a study on the compatibility between swampland conjectures and asymptotically safe gravity.}. The resulting effective action reads
\begin{eqaed}\label{eq:effactna}
    \Gamma_\mathrm{string} =\frac{ \text{Vol}_d}{16\pi G_\mathrm{N}} \int dt \, \, n \, e^{-\Phi} \left[ -\,  \frac{\dot{\Phi}^2}{n^2} + \left(\Lambda+\tilde{c}\,\sqrt{\frac{H^{2}}{n^2 \Lambda}}\right) \right] \, , 
\end{eqaed}
where we have restored the lapse $n$, we have introduced the effective coupling $\tilde{c}=\frac{\sqrt{\Lambda}}{16\pi G_\mathrm{N}}c_1$, and we have identified $c_0\equiv\frac{\Lambda}{16\pi G_\mathrm{N}}$, where $\Lambda\approx 1/\alpha'$ is the (leading contribution to the) string-frame cosmological constant\footnote{More precisely, $\Lambda \propto \frac{D_\text{crit}-D}{\alpha'}$ encodes the deviation from the critical dimension, namely $D_\text{crit} = 26 \, ,  \, 10$ for bosonic strings and superstrings respectively~\cite{Veneziano:1991ek, Fradkin:1984pq, Callan:1985ia}.}. The corresponding cosmological field equations take the simple form
\begin{eqaed}\label{eq:cosmoeq_1}
\dot{\Phi}^2=-\Lambda \,,\qquad \ddot{\Phi}=\frac{\tilde{c}}{2}\,\sqrt{\frac{H^2}{\Lambda}} \,,\qquad \frac{\tilde{c}}{\Lambda^2}\,\frac{|H|}{H} \, \dot{\Phi}=0 \,,\qquad
\end{eqaed}
and it is straighforward to see that if $\tilde{c}\neq0$ no dS solution with $H = \text{const.}$ is allowed, not even with a time-varying dilaton~$\phi$. If instead $\tilde{c}=0$, the above cosmological equations admit a unique solution with constant $H$ and
\begin{eqaed}
\Phi(t)=C\pm\sqrt{\Lambda}\,t\,,
\end{eqaed}
$C$ being an integration constant. However, this quasi-dS solution is only realized in an unphysical regime, since $\tilde{c}=0$ entails trivial gravitational dynamics at the level of the effective action $\Gamma_\mathrm{string}$. Let us remark that the solution above, while exact, might not be unique. A proper analysis of the flow equations, including the space of initial conditions compatible with UV fixed points, appears prohibitive in general, but, as we shall now discuss, it is feasible in $2+\epsilon$ dimensions.

The effective action of Eq.~\eqref{eq:effactna} is neither analytic in the curvatures (thus it is unable to reproduce string theory in the UV or general relativity in the IR) nor expected to be unique. We are therefore led to seek analytic, and possibly \textit{general}, solutions to the RG equations. To this end, it is useful to employ an $\epsilon$-expansion of the effective action about $D = 2$, along the lines of~\cite{1979W,Gastmans:1977ad,Christensen:1978sc,Kawai:1989yh,Kawai:1992np,Kawai:1993mb,Kawai:1995ju,Aida:1996zn}  and then try to extend the results to $D=4$.
The rationale behind this approach rests on the observation that the Newton coupling is classically marginal in $D=2$ dimensions, and therefore the RG flow dramatically simplifies expanding $D = 2+\epsilon$ for $\epsilon \ll 1$. \commie{In Einstein gravity this procedure is already instructive~\cite{1979W,Gastmans:1977ad,Christensen:1978sc,Kawai:1989yh,Kawai:1992np,Kawai:1993mb,Kawai:1995ju,Aida:1996zn}, but in order to derive all-order corrections it is necessary to include the stringent constraints of T-duality in mini-superspace.} Since one expects a UV fixed point for $g_k$ of order $\mathcal{O}(\epsilon)$, our starting point is the ansatz $g_k = \epsilon \, \gamma_k$, with $\gamma_k = \mathcal{O}(1)$, while the ansatz
\begin{eqaed}
	F_k(H^2) = \frac{v_k(H^2)}{\epsilon\, \gamma_k} + w_k(H^2)\,\,,
\end{eqaed}
with $v_k(H^2) \, , \, w_k(H^2) = \mathcal{O}(1)$, is motivated by the expected IR behavior of $F_k$, whose corrections ought to be sub-leading in $\epsilon$. The resulting flow equations simplify dramatically at leading order in $\epsilon$. As a result, the dimensionless Newton coupling flows according to $\gamma_k = \gamma_* (1 + c \, k^{-\epsilon})^{-1}$, with $c$ a constant. This flow mirrors that of Eq.~\eqref{eq:runningg} for $d=1+\epsilon$, and in particular it starts from $\gamma_* = \frac{3}{2} \, \pi^2$ in the UV and ends at zero the IR for $c > 0$, where $\gamma_k \sim \frac{\gamma_*}{c} \, k^{\epsilon}$. The corresponding IR value of the dimensionful Newton coupling is then $G_k \sim \frac{3\pi}{32} \, \frac{\epsilon}{c}$. The next step involves the sub-leading corrections, where $\dot{\gamma_k} \sim \epsilon \, \gamma_k$ can be neglected with respect to the other contributions. The sub-leading flow equations can be solved in closed form, and in the IR one obtains higher-derivative corrections to the low-energy effective action to all orders in $H$, and thus in $\alpha'$, since the $k$ dependence correctly disappears.

\begin{figure}
\hspace{-0.4cm}\includegraphics[scale=0.45]{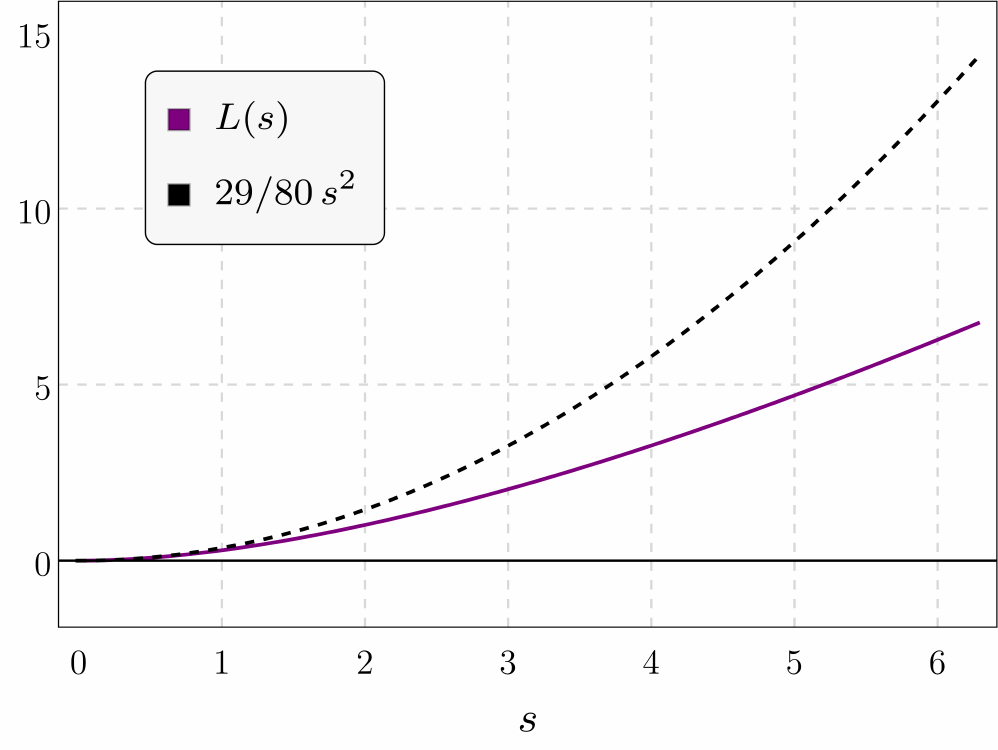}
\caption{The function $L(s)$ (purple curve) encoding higher-derivative $\alpha'$ corrections to the classical action. As expected, terms proportional to $s^2\sim R^2$ (black dashed line) dominate the $\alpha'$-expansion of the effective action at low curvatures. \label{Fig:efflag}}
\end{figure}

The \textit{most general analytic solution} to the exact flow equation in $2+\epsilon$ dimensions compatible with a UV fixed point leads to the $\mathcal{O}(\epsilon)$ effective action~\cite{Basile:2021amb}
\begin{eqaed}\label{eq:string-frame_IR_EA}
	&\Gamma_{\text{string}} = \frac{\text{Vol}_{1+\epsilon}}{16\pi G_{\text{N}}} \int dt \, n \, e^{-\Phi} \times \\ &\left[\Lambda - \frac{\dot{\Phi}^2}{n^2}+ \frac{H^2}{n^2} + \frac{8 G_{\text{N}} \Lambda}{3\pi} \, L\left(\frac{H^2}{n^2\,\Lambda} \right) \right] \, ,
\end{eqaed}
where the relevant deformation $\Lambda$ from the UV fixed point controls the validity of the low-energy expansion, and
\begin{eqaed}\label{eq:L_func}
    L(s) & \equiv - 1 - \frac{23}{12} \, s + \left(\frac{3}{2} + s \right) \log\left(1 + \frac{s}{2} \right) \\
    & + \left(1+s\right)^{\frac{3}{2}} \, \sqrt{\frac{2}{s}} \, \text{arctanh}\left( \sqrt{\frac{s}{2\left(1+s\right)}} \right)
\end{eqaed}
is depicted in Fig.~\ref{Fig:efflag}. As expected, quadratic corrections to the classical action dominate at low curvatures ($H\ll\Lambda$).
The effective cosmological equations in $(2+\epsilon)$-dimensions read
\begin{eqaed}\label{eq:cosmoeq_2}
& \dot{\Phi}^2=H^2\left[1+\frac{16G}{3\pi}L'\left(\frac{H^2}{\Lambda}\right)\right]-\Lambda\left[1+\frac{8G}{3\pi}L\left(\frac{H^2}{\Lambda}\right)\right] \,,\\
& \ddot{\Phi}=H^2\left[1+\frac{8G}{3\pi}L'\left(\frac{H^2}{\Lambda}\right)\right] \,,\\
& \frac{\dot{H}}{H}=\dot{\Phi}-\frac{8 G}{3 \pi }L'\left(\frac{H^2}{\Lambda}\right)\,\left(\frac{\dot{H}}{H}-\dot{\Phi} \right)-\frac{16 G H \dot{H}}{3 \pi  \Lambda }L''\left(\frac{H^2}{\Lambda }\right) \,,
\end{eqaed}
where we have omitted the dependence on $t$ for brevity. Specializing  Eq.~\eqref{eq:cosmoeq_2} to the case of constant $H$ yields the existence conditions for non-perturbative dS solutions  discussed in~\cite{Hohm:2019ccp, Nunez:2020hxx}. It is straightforward to see that dS solutions are not allowed by Eq.~\eqref{eq:cosmoeq_2}, since the conditions of~\cite{Hohm:2019ccp} cannot be met by the all-order $\alpha'$-corrections encoded in the (monotone increasing) $L(s)$ of Eq.~\eqref{eq:L_func}. Even in the Einstein-frame, the Hubble parameter satisfies~\cite{Hohm:2019ccp}
\begin{eqaed}\label{eq:einstein_hubble}
H_E \overset{\epsilon \to 0}{\sim} \frac{d}{dt} \, e^{\frac{1}{\epsilon} (\sigma + \Phi)} \, ,
\end{eqaed}
which entails $H \sim - \, \dot{\Phi}$ for an Einstein-frame dS solution. This is also non-trivially incompatible with Eq.~\eqref{eq:cosmoeq_2}. Even in the Einstein-frame, the Hubble parameter satisfies~\cite{Hohm:2019ccp}
\begin{eqaed}\label{eq:einstein_hubble}
	H_E \overset{\epsilon \to 0}{\sim} \frac{d}{dt} \, e^{\frac{1}{\epsilon} (\sigma + \Phi)} \, ,
\end{eqaed}
which entails $H \sim - \, \dot{\Phi}$ for an Einstein-frame dS solution. This is also non-trivially incompatible with Eq.~\eqref{eq:cosmoeq_2}. Analogously, the non-analytic effective action in Eq.~\eqref{eq:effactna} does not allow dS solutions. Moreover, at least within polynomial truncations up to order $\mathcal{O}(H^{10})$, the effective action in~Eq.\eqref{eq:string-frame_IR_EA} does not extend to~$D=4$~\cite{Basile:2021amb}. We are thus led to conclude that, in this setting, insofar as only mini-superspace cosmologies are included in the gravitational functional integral, our results point to a tension between dS solutions and curvature corrections in string theory. Our findings thus support the no-dS swampland conjecture~\cite{Obied:2018sgi, Ooguri:2018wrx, Garg:2018reu} in the presence of all-order effects in~$\alpha'$.

Let us finally remark that, although a highly-curved dS spacetime cannot be described at the level of two-derivative (super)gravity, it could have in principle arisen from our infinite-derivative field equations. In this case, the resulting cosmological constant would have most likely been of the order of the string scale.


\textit{Conclusions.}---
Understanding higher-derivative corrections and all-order effects appears of fundamental importance for high-energy regimes of gravity. In particular, they could play a crucial role in the resolution of singularities and in the structure of the fundamental degrees of freedom and their interactions. Moreover, they could be crucial to assess the existence of de Sitter (dS) solutions, in particular in the context of string theory. 
In this respect, while string theory provides, at least in some regimes, well-defined algorithms to systematically compute such corrections, earlier attempts to derive them in the general case were met by intricate technical difficulties. Therefore, we are compelled to investigate novel directions in order to shed light on these relevant matters.

To this end, symmetries play a key role. Indeed, although direct computations of all-order effects are extremely involved in general, stringy symmetries such as T-duality constrain their form substantially~\cite{Meissner:1991zj, Meissner:1996sa, Hohm:2015doa, Hohm:2019ccp, Hohm:2019jgu}, thus allowing to investigate the existence of dS solutions in string theory beyond tree-level. Earlier works in this direction~\cite{Boulware:1986dr, Bento:1995qc, Maroto:1997aw} involved the leading curvature corrections to the classical action and discovered that  -- \textit{to leading order in $\alpha'$} -- dS vacua are excluded by the simultaneous presence of the dilaton and $\alpha'$-corrections~\cite{Boulware:1986dr, Bento:1995qc}.

In this work we have taken a major step forward by resumming \textit{all} $\alpha'$ corrections on cosmological backgrounds. Specifically, combining the constraints on stringy cosmological effective actions 
with FRG techniques~\cite{Basile:2021amb}, in this Letter we have derived the cosmological field equations associated with two closed-form $\alpha'$-corrected effective actions and we have investigated the existence of dS solutions in the presence of all-order $\alpha'$-corrections. No dS solution seems to be allowed. Our findings  thus support the implications of the relevant swampland conjectures~\cite{Obied:2018sgi, Ooguri:2018wrx, Garg:2018reu} in the presence of curvature corrections to all orders in~$\alpha'$. 

All in all, within the framework of string theory, our results appear to point at one (or more) of the following possibilities:
\begin{itemize}
\item The existence of dS solutions in string theory requires inhomogeneities and/or anisotropies, e.g., to generalize the mini-superspace ansatz to Bianchi-like models
\item dS solutions require the inclusion of string-loop corrections
\item No dS solution is allowed in string theory, and early/late-time phases of accelerated expansion are driven by quintessence (or analogous scenarios)~\cite{Agrawal:2019dlm, Banerjee:2020xcn, DiValentino:2021izs}.
\end{itemize}

Comparing the FRG approach presented in this work with the results of perturbative string theory in its critical dimension remains an important, if daunting, undertaking. As pointed out in~\cite{Basile:2021amb}, it is currently challenging to test our formalism due to the considerable difficulty of solving the FRG equations in sufficient generality in higher dimensions. The only setting in which this is currently feasible is within an epsilon-expansion about $D=2$, where we managed to obtain the most general flow compatible with a UV fixed point. Comparing our results to \emph{bona fide} string perturbation theory computations would require, at least, resumming this expansion or finding general UV-complete surfaces in higher dimensions. Alternatively, one would need to compute the $\alpha'$-exact effective action in non-critical dimensions in string theory, which is currently out of reach (except for special -- non-cosmological -- backgrounds).
	
At the same time, while difficult to test, the interplay between the FRG and the symmetries of string theory could open new doors, allowing to investigate important non-perturbative aspects of phenomenologically relevant scenarios otherwise unaccessible with standard perturbative methods.
In particular, considerations similar to those that we have put forth in this Letter, as well as in~\cite{Basile:2021amb}, could be applied to a number of scenarios where all-order effects are expected to be important, for instance the cancellation of Weyl anomalies of the string worldsheet~\cite{worldsheetFRG}, thereby providing a new and rich avenue of research. We would like to explore these intriguing ideas and scenarios in future work.

\textit{Acknowledgments.}---
{The authors would like to thank A. Bonanno, A. Eichhorn, B. Knorr, A. Sagnotti and F. Saueressig for discussions. A.P. acknowledges support by Perimeter Institute for Theoretical Physics. Research at Perimeter Institute is supported in part by the Government of Canada through the Department of Innovation, Science and Economic Development Canada and by the Province of Ontario through the Ministry of Colleges and Universities. The work of I.B.\ was supported by the Fonds de la Recherche Scientifique - FNRS under Grants No.\ F.4503.20 (``HighSpinSymm'') and T.0022.19 (``Fundamental issues in extended gravitational theories'').}






\bibliography{t-dual.bib}

\end{document}